\documentstyle[sprocl]{article}
\def\Journal#1#2#3#4{{#1} {\bf #2}, #3 (#4)}
\def\CQG{\em Class. Quantum Gravity}
\def\JHEP{\em J. High Energy Phys.}

\def\NPB{{\em Nucl. Phys.} B}
\def\PLB{{\em Phys. Lett.} B}
\def\PRL{\em Phys. Rev. Lett.}
\def\PRD{{\em Phys. Rev.} D}

\def\be{\begin{equation}}
\def\ee{\end{equation}}
\def\ba{\begin{eqnarray}}
\def\ea{\end{eqnarray}}

\def\a{\alpha}
\def\b{\beta}
\def\g{\gamma}

\def\e{\epsilon}

\def\l{\lambda}
\def\L{\Lambda}
\def\m{\mu}
\def\n{\nu}

\def\r{\rho}
\def\s{\sigma}

\def\w{\omega}

\def\de{\partial}
\begin{document}
\pagestyle{empty}
\hbox{\hskip 12cm ROM2F-98/42  \hfil}
\vskip 1cm
\begin{center}
{\Large  \bf  Self-dual \ Tensors \ in \vskip 24pt
 Six-Dimensional \ Supergravity}

\vspace{2cm}

{\large \large  Fabio Riccioni \ \ and \ \ Augusto Sagnotti}

\vspace{1cm}

{\sl Dipartimento di Fisica\\
Universit{\`a} di Roma \ ``Tor Vergata'' \\
I.N.F.N.\ - \ Sezione di Roma \ ``Tor Vergata'' \\
Via della Ricerca Scientifica, 1 \\
00133 \ Roma \ \ ITALY}
\vskip 1cm
{\bf Abstract}
\end{center}
\vskip 18pt
{We review some properties of the field equations of six-dimensional
(1,0) supergravity coupled to tensor and vector multiplets, and in
particular their relation to covariant and consistent anomalies and
a peculiar Noether identity for the energy-momentum tensor. We also
describe a lagrangian formulation for this system,
obtained applying the Pasti-Sorokin-Tonin prescription.}

\vskip 1.5cm
\begin{center}
{Based on talks presented at the 1998 Trieste Conference on
``Superfivebranes and Physics in 5+1 Dimensions", at the 1998 DESY Workshop
on ``Conformal Field Theory of D branes" and at the 1998 Como-Centro Volta
RFBF-INTAS School on ``Advances in Quantum Field Theory, 
Statistical Mechanics and Dynamical Systems"} 
\end{center}
\vfill\eject                    
\pagestyle{plain}
\title{Self-dual Tensors in Six-Dimensional Supergravity}

\author{Fabio Riccioni \ and \ Augusto Sagnotti}
\address{
Dipartimento di Fisica, Universit\`a di Roma ``Tor Vergata''\\
I.N.F.N., Sezione di Roma ``Tor Vergata''\\
Via della Ricerca Scientifica, 1 \ 00133 Roma \ ITALY}
\maketitle
\abstracts{
We review some properties of the field equations of six-dimensional
(1,0) supergravity coupled to tensor and vector multiplets, and in
particular their relation to covariant and consistent anomalies and
a peculiar Noether identity for the energy-momentum tensor. We also
describe a lagrangian formulation for this system,
obtained applying the Pasti-Sorokin-Tonin prescription.}

\section{Introduction}

Six-dimensional $(1,0)$ supergravity is built out of four types of
multiplets~\cite{ns,romans}.  Aside from the gravitational multiplet $(e_\mu^a,
\psi_{L,\mu}, B^+_{\mu\nu})$, one has the option to add
tensor multiplets $(B^-_{\mu\nu},
\chi_R,\varphi)$, vector multiplets $(A_\mu,\lambda_L)$ and hypermultiplets.
This talk is devoted to some notable properties of the low-energy couplings
between the first three types of multiplets, and is meant to complement
our previous short review~\cite{trieste97}. Couplings between several tensor
multiplets have entered the stage relatively recently, since  
they emerge rather naturally only 
in perturbative type-I vacua~\cite{tensormults}. On the
other hand, perturbative heterotic vacua
always involve a single tensor multiplet, that is responsible for the
corresponding Green-Schwarz mechanism~\cite{gs}. 

The couplings
between vector and tensor multiplets are rather unconventional, since they
are ``classically anomalous'' and conspire to
implement a generalized Green-Schwarz mechanism~\cite{as}.  
Limitedly to the gauge anomalies, in six dimensions
the Green-Schwarz mechanism is visible in the
low-energy field equations, and brings about a number of oddities,
most notably some singularities in the moduli space of tensor 
multiplets~\cite{as}.  These signal an important new phenomenon, a phase
transition~\cite{dmw} related to strings of vanishing 
tension, likely to be of quite some interest in
the coming years.

In formulating the low-energy couplings between tensor and vector multiplets,
one has two natural options.  The first is related to covariant field equations
and to the corresponding covariant anomalies~\cite{as,rs}. It has the virtue of 
respecting gauge covariance and supersymmetry, but the resulting 
field equations are not integrable.  The second
is related to consistent, and thus integrable, field equations~\cite{fms,frs}. 
These may be derived from an action principle that satisfies Wess-Zumino 
consistency conditions, and as a result embody a supersymmetry anomaly.  The
complete field equations, first obtained in this framework~\cite{frs}, are
not unique \footnote{A number of fermionic couplings, also to
hypermultiplets, with additional 
singularities related to those of ref. \cite{as}, were anticipated in 
ref. \cite{ns2}.}. A quartic coupling
for the gauginos, proportional to ${(\bar{\lambda} \gamma^\mu \lambda})^2$,
is undetermined, while the gauge algebra contains an
extension that makes the construction consistent for any choice of it. 
We would like to stress that all
these features are determined by local couplings
in these ``classical'' field equations, that are thus a remarkable laboratory
for current algebra. Indeed, a closer scrutiny of their properties~\cite{rs}
brought about one further surprise: in a theory with gauge and supersymmetry
anomalies, gravitational anomalies are {\it not} directly related to the
divergence of the energy-momentum tensor. Thus, in a theory without
gravitational anomalies the energy-momentum tensor need not be divergenceless. 

We shall return to the energy-momentum tensor in the next Section,
where we shall also review our recent work on the completion of the covariant
equations~\cite{rs}. 
Aside from this general lesson, there are also a number of subtleties related to the 
rigid limit of the coupled equations. These are not considered here, but have been
discussed at length in ref. \cite{rigid}.
Actually, in mentioning an action principle, we have
been somewhat cavalier about the presence of (anti)self-dual antisymmetric 
tensors, that bring about a number of traditional difficulties. The long-standing 
problem of giving a covariant action principle to this type of fields has
recently been given a compact solution, in terms of a single auxiliary scalar
mode, by Pasti, Sorokin and Tonin (PST)~\cite{pst}. The last Section is 
devoted to the completion of the consistent action principle of ref.~\cite{frs}
using their method. This new result, not contained in the original 
presentation, is included here for completeness.

\section{Consistent and covariant field equations and their anomalies}

In this Section we would like to review some basic properties 
of six-dimensional $(1,0)$ supergravity coupled to vector and tensor
multiplets. This model is rather peculiar, since it embodies gauge
anomalies induced by tensor couplings~\cite{as,fms,frs,rs}, to be disposed
of by fermion loops that one is actually not accounting for. 
Consequently, in formulating the field equations, one is faced
with the familiar choice between covariant and consistent anomalies.
The former originate from a set of covariant non-integrable equations 
invariant under local
supersymmetry, while the latter, generally regarded as the most interesting ones, 
originate from a set of integrable equations, with a consistent 
supersymmetry anomaly 
related to the consistent gauge anomaly by Wess-Zumino consistency conditions.

To lowest order in the fermi fields, in the conventions of ref. \cite{frs},
the fermionic equations are~\cite{as,fms}
\ba
& & \g^{\m\n\r} D_\n \Psi_\r + v_r H^{r \m\n\r} \g_\n \Psi_\r
-\frac{i}{2} x^m_r H^{ r \m\n\r} \g_{\n\r} \chi^m  
\nonumber\\
& &+ \frac{i}{2} x^m_r \de_\n v^r \g^\n \g^\m \chi^m -\frac{1}{\sqrt{2}}
v_r c^{rz} tr_z (F_{\sigma \tau} \g^{\sigma \tau} \g^\m \l )=0 \quad ,
\label{gaugino}
\ea
\ba
& & \g^\m D_\m \chi^m -\frac{1}{12} v_r H^{r \m\n\r} 
\g_{\m\n\r} \chi^m
-\frac{i}{2} x^m_r H^{r \m\n\r} \g_{\m\n} \Psi_\r \nonumber\\
& & - \frac{i}{2} x^m_r \de_\n v^r \g^\m \g^\n \Psi_\m -\frac{i}{\sqrt{2}}
x^m_r c^{rz} tr_z (F_{\m\n}\g^{\m\n} \l ) =0 
\ea
and
\ba
& & v_r c^{rz}\g^\m D_\m \l +\frac{1}{2} (\de_\m v_r ) c^{rz} \g^\m \l
+\frac{1}{2\sqrt{2}} v_r c^{rz} F_{\a\b} \g^\m \g^{\a\b} \Psi_\m \nonumber\\
& &  +\frac{i}{2\sqrt{2}}x^m_r c^{rz} F_{\m\n} \g^{\m\n} 
\chi^m -\frac{1}{12} c^{rz} H_{r \m\n\r} \g^{\m\n\r} \l =0 \quad ,
\ea
while the consistent bosonic equations are~\cite{fms}
\ba
& & R_{\m\n} -\frac{1}{2} g_{\m\n} R + \de_\m v^r \de_\n v_r 
-\frac{1}{2} g_{\m\n} \de_\a v^r \de^\a v_r -
G_{rs} H^r_{\m\a\b} H^s{}_\n{}^{\a\b} \nonumber\\
& &+ 4 v_r c^{rz}
tr_z (F_{\a\m} F^\a{}_\n -\frac{1}{4} g_{\m\n} F_{\a\b}F^{\a\b})=0\quad ,
\ea
\be
x^m_r D_\m (\de^\m v^r ) +\frac{2}{3} x^m_r v_s H^r_{\a\b\g} 
H^{s \a\b\g} -x^m_r c^{rz} tr_z (F_{\a\b} F^{\a\b})=0 \label{vectorcov}
\ee
and
\ba
& & D_\m (v_r c^{rz} F^{\m\n} ) -  c^{rz} G_{rs} H^{s \n\r\sigma}
F_{\r\sigma}   - \frac{1}{8e} \e^{\n\r\a\b\g\delta} c_r^z A_\r c^{rz^\prime} 
tr_{z^\prime} (F_{\a\b}F_{\g\delta}) \nonumber\\
& & -\frac{1}{12e}\e^{\n\r\a\b\g\delta} c_r^z F_{\r\a} c^{rz^\prime}
\w^{z^\prime}_{\b\g\delta} = 0 \quad. \label{newvectoreq}
\ea
Moreover, the tensor fields satisfy (anti)self-duality conditions, conveniently
summarized as~\cite{fms}
\be
G_{rs} H^{s \m\n\r} =\frac{1}{6e} \e^{\m\n\r\a\b\g} H_{r \a\b\g}\quad ,
\label{selfdual}
\ee
where $G_{rs} =v_r v_s + x^m_r x^m_s$. 

In this talk, as in refs.~\cite{as,fms,frs,rs}, 
we are confining our attention to
residual anomalies, that correspond to reducible traces. In type-I vacua,
these are the contributions left over by tadpole conditions. In String
Theory, all residual anomalies are absent {\it a priori}, since they
would draw their origin from the non-planar one-loop amplitude, that 
is regulated by the momentum flow. On the other hand, in Field Theory 
the residual anomalies draw their origin from fermion loops, and are
disposed of by the Green-Schwarz couplings that we are actually discussing.
In particular, eqs.  (\ref{gaugino})-(\ref{newvectoreq}) embody a 
reducible gauge anomaly~\cite{fms}
\be  {\cal{A}}_\L =- \frac{1}{4} \e^{\m\n\a\b\g\delta} c_r^z c^{rz^\prime} tr_z
(\L
\de_\m A_\n ) tr_{z^\prime} (F_{\a\b} F_{\g\delta} )\quad ,
\label{consanomaly}
\ee 
and a corresponding supersymmetry anomaly
\be
{\cal{A}}_\e = - c^{rz} c_r^{z^\prime } [
tr_z (\delta_\e A \ A ) tr_{z^\prime} (F^2 )
-2 tr_z (\delta_\e A \ F )
\w^{z^\prime}_3 ]\quad ,\label{susyan}
\ee
determined by Wess-Zumino consistency conditions~\cite{wz}. 

The complete field equations were obtained~\cite{frs} from the
commutator of two supersymmetry transformations on the fermi fields, in
the spirit of refs.~\cite{schwarz,romans}. However, in this case
one is actually solving Wess-Zumino consistency conditions, and these
do not fix a cubic contribution to the gaugino equation
and a related quartic contribution to the vielbein equation. This lack of
uniqueness, a rather surprising phenomenon in supergravity, reflects a
familiar property of anomalies, that are defined up to the variation
of a local functional. It would be interesting to investigate further this issue
directly in String Theory.

The gauge anomaly ${\cal A}_\L = \delta_\L {\cal L}$ naturally satisfies the
condition
\be {\cal A}_\L= -tr (\L D_\m J^\m )\quad, \label{div1}
\ee where $J^\m =0$ is the complete field equation of the vector field. One can
similarly show that the supersymmetry anomaly is related to the field equation
of the gravitino, that we write succinctly
$\tilde{J}^\m =0$, according to
\be {\cal A}_\e = -(\bar{\e} D_\m \tilde{J}^\m ) \quad .\label{div2}
\ee

Moreover, these equations embody an amusing Noether identity for 
the energy-momentum tensor, a general result in Field Theory, that came as
a little surprise to us. In a theory with gauge and
supersymmetry anomalies, the gravitational anomaly is {\it not} simply
related to the divergence of the energy-momentum tensor~\cite{rs}, since
\be 
{\cal A}_{\xi} = \delta_{\xi}{\cal L}= 2 \xi_\n D_\m T^{\m\n} +
\xi_\n tr (A^\n D_\m J^\m ) + \xi_\n  (\bar{\Psi}^\n D_\m
\tilde{J}^\m )\quad. \label{emtnoether}
\ee 
In particular, in our case we are not accounting for gravitational
anomalies, that would  result in higher-derivative couplings, and indeed one can
verify that the divergence of the energy-momentum tensor does not vanish, but
satisfies the relation
\be
D_\m T^{\m\n} =-\frac{1}{2} tr(A^\n D_\m J^\m )- \frac{1}{2} 
(\bar{\Psi}^\n D_\m \tilde{J}^\m )\quad. \label{emtdiv}
\ee
Thus, the lack of conservation of matter currents feeds an algebraic
inconsistency in the Einstein equations. We are not aware of a previous 
discussion of this simple but neat result in the literature.

Consistent and covariant gauge anomalies are related  by the divergence of
a local functional~\cite{bz}. In six dimensions, to lowest order in the
fermi fields,
the residual covariant  gauge anomaly is~\cite{as}
\be {\cal A}_\L^{cov}=\frac{1}{2} \e^{\m\n\a\b\g\delta} c^{rz}c_r^{z^\prime}tr_z
(\L F_{\m\n}) tr_{z^\prime} (F_{\a\b} F_{\g\delta} ) \quad , \label{covanomaly}
\ee and is related to the consistent anomaly according to
\be {\cal A}_\L^{cons} +tr [\L D_\m f^\m ] ={\cal A}_\L^{cov}\quad
,\label{covcons}
\ee where
\be f^\m = - c_r^z c^{r z^\prime} \e^{\m\n\a\b\g\delta} \lbrace \frac{1}{4}
 A_\n tr_{z^\prime}( F^\prime_{\a\b} F^\prime_{\g\delta} )
+ \frac{1}{6}   F_{\n\a} 
\ \w^\prime_{\b\g\delta} \rbrace \quad .\label{fmu}
\ee  Comparing eq. (\ref{fmu}) and eq. (\ref{susyan}), one can see that
\be {\cal A}_\e = tr(\delta_\e A_\m \ f^\m )  \quad,
\ee and this implies that the transition from consistent  to  covariant anomalies
turns a model with a supersymmetry anomaly into one without any~\cite{as,fms}. Indeed, six-dimensional supergravity coupled to vector and tensor
multiplets was originally formulated in this fashion in ref.~\cite{as} to lowest
order in the fermi fields, extending the work of Romans~\cite{romans}.

This observation can be generalized to all
orders in the fermi fields. The complete supersymmetry anomaly has the form
\be 
{\cal A}_\e = tr(\delta_\e A_\m \ f^\m ) 
+\delta_\e e_\m{}^a \ g^\m{}_a \quad ,
\ee 
where
\ba  f^\m &=&c_r^z c^{r z^\prime} tr_{z^\prime} \lbrace -\frac{1}{4}
\e^{\m\n\a\b\g\delta} A_\n F^\prime_{\a\b} F^\prime_{\g\delta}
-\frac{1}{6}
\e^{\m\n\a\b\g\delta} F_{\n\a} 
\w^\prime_{\b\g\delta} \nonumber\\ & + & \frac{i e}{2} F_{\n\r}
(\bar{\l}^\prime
\g^{\m\n\r} 
\l^\prime )+\frac{i e}{2} (\bar{\l} \g^{\m\n\r} \l^\prime )
F^\prime_{\n\r}  + ie (\bar{\l}\g_\n \l^\prime ) F^{\prime \m\n}
\nonumber\\ & - & 
 \frac{e}{2\sqrt{2}} (\bar{\l} \g^\m
\g^\n \g^\r 
\l^\prime )(\bar{\l}^\prime \g_\n \Psi_\r )   +  \frac{e x^m_s
c^{s z^\prime}}{v_t c^{t z^\prime}} [-\frac{3 i}{2\sqrt{2}}
 (\bar{\l} \g^\m \l^\prime )(\bar{\l}^\prime \chi^m ) \nonumber\\
& - & \frac{i}{4 \sqrt{2}} (\bar{\l} \g^{\m\n\r} \l^\prime
)(\bar{\l}^\prime
\g_{\n\r}\chi^m )  -  \frac{i}{2\sqrt{2}} 
(\bar{\l} \g_\n
\l^\prime )(\bar{\l}^\prime \g^{\m\n} \chi^m )] \nonumber\\
& +& e [-i\a
\hat{F}_{\n\r}(\bar{\l}^\prime \g^{\m\n\r}\l^\prime )+   i\a(\bar{\l}
\g^{\m\n\r}\l^\prime ) 
\hat{F}^\prime_{\n\r} - 6i\a(\bar{\l}\g^\n 
\l^\prime ) \hat{F}^\prime_{\m\n}]\nonumber\\ 
& + &  
\frac{e x^m_s c^{s z^\prime}}{v_t c^{t z^\prime}} [-i \a \sqrt{2}
(\bar{\l}\g^\m \l^\prime ) (\bar{\l}^\prime \chi^m )
+\frac{i\a}{2\sqrt{2}}(\bar{\l}\g_{\n\r}\chi^m )(\bar{\l}^\prime 
\g^{\m\n\r}\l^\prime )]\nonumber\\ 
& + &  \frac{e x^m_s c^{s
z}}{v_t c^{t z}}  [\frac{i \a}{\sqrt{2}}(\bar{\l}\g^\m \l^\prime
)(\bar{\l}^\prime
\chi^m ) -\frac{i \a}{2\sqrt{2}}(\bar{\l} \g^{\m\n\r} \l^\prime )(\bar{\l}^\prime
\g_{\n\r} \chi^m )\nonumber\\ & + & \frac{i \a}{\sqrt{2}}(\bar{\l}\g_\n \l^\prime
)(\bar{\l}^\prime \g^{\m\n}
\chi^m ) ]
\rbrace 
\label{fmutot}
\ea
differs from the expression in eq. (\ref{fmu}) by the addition of higher-order
fermionic terms, while
\be
g^\m{}_a = c_r^z \ c^{r z^\prime} tr_{z,z^\prime} \lbrace\frac{e}{32} ( \bar{\l} \g^{\m\n\r} \l)(\bar{\l}^\prime \g_{a\n\r} 
\l^\prime) + \frac{\a e e^\m{}_a}{2} ( \bar{\l} \g^{\n} \l^\prime)(\bar{\l} \g_{\n} 
\l^\prime) \rbrace \quad . \label{gma}
\ee 
It should be appreciated that both expressions depend on the {\it arbitrary} parameter
$\alpha$. 
Modifying the vector equation so that
\be 
(eq. \ A^\m )_{(cov)} \equiv J^\m_{(cov)}=\frac{\delta {\cal L}}{\delta A_\m } -
f^\m \quad ,
\label{modifiedvector}
\ee 
and similarly for the Einstein equation, the resulting theory is
supersymmetric but no longer integrable. 
The divergence of the complete covariant vector equation satisfies
\be tr (\L D_\m J^\m_{(cov)} )=-{\cal A}_\L^{cov}\quad ,
\ee 
where 
\ba
{\cal{A}}_\L^{cov} & = & c^{rz}c_r^{z^\prime}tr_{z,z^\prime}\lbrace
\frac{1}{2} \e^{\m\n\a\b\g\delta} (\L F_{\m\n})  (F^\prime_{\a\b}
F^\prime_{\g\delta} )  +i e \L F_{\n\r} (\bar{\l}^\prime
\g^{\m\n\r}D_\m 
\l^\prime )
\nonumber\\  
& + & \frac{i e}{2} \L D_\m (\bar{\l} \g^{\m\n\r} \l^\prime )
F^\prime_{\n\r}  + e \L D_\m \lbrace i(\bar{\l}\g_\n \l^\prime ) F^{\prime \m\n}
-  i\a
\hat{F}_{\n\r}(\bar{\l}^\prime \g^{\m\n\r}\l^\prime )\nonumber\\
& + &  i\a(\bar{\l}
\g^{\m\n\r}\l^\prime ) 
\hat{F}^\prime_{\n\r} - 6i\a(\bar{\l}\g_\n 
\l^\prime ) \hat{F}^{\prime \m\n}
- \frac{1}{2\sqrt{2}} (\bar{\l} \g^\m \g^\n
\g^\r 
\l^\prime )(\bar{\l}^\prime \g_\n \Psi_\r )\nonumber\\ 
& + &  \frac{ x^m_s c^{s z^\prime}}{v_t c^{t z^\prime}} [-\frac{3 i}{2\sqrt{2}}
(\bar{\l} \g^\m \l^\prime )(\bar{\l}^\prime \chi^m )
- \frac{i}{4 \sqrt{2}} (\bar{\l} \g^{\m\n\r} \l^\prime )(\bar{\l}^\prime
\g_{\n\r}\chi^m ) \nonumber\\
& - & \frac{i}{2\sqrt{2}}  (\bar{\l} \g_\n
\l^\prime )(\bar{\l}^\prime \g^{\m\n} \chi^m ) + 
\frac{i\a}{2\sqrt{2}}(\bar{\l}\g_{\n\r}\chi^m )(\bar{\l}^\prime 
\g^{\m\n\r}\l^\prime )\nonumber \\
& - & i \a \sqrt{2}
(\bar{\l}\g^\m \l^\prime ) (\bar{\l}^\prime \chi^m ) ]
+ \frac{x^m_s c^{s
z}}{v_t c^{t z}}  [\frac{i \a}{\sqrt{2}}(\bar{\l}\g^\m \l^\prime
)(\bar{\l}^\prime
\chi^m ) \nonumber\\
& - & \frac{i \a}{2\sqrt{2}}(\bar{\l} \g^{\m\n\r} \l^\prime )(\bar{\l}^\prime
\g_{\n\r} \chi^m ) +\frac{i \a}{\sqrt{2}}(\bar{\l}\g_\n \l^\prime
)(\bar{\l}^\prime \g^{\m\n}
\chi^m ) ]\rbrace
\rbrace \quad 
\label{completecovanomaly}
\ea
now contains higher-order fermi terms~\cite{rs}.

Finally, one can study the divergence of the Rarita-Schwinger and Einstein 
equations in the covariant model. With the covariant
equations obtained from the consistent ones by the redefinition of eq. 
(\ref{modifiedvector}) and by
\be 
(eq. \ e^\m{}_a )_{(cov)}=\frac{\delta {\cal L}}{\delta e_\m{}^a } -g^\m{}_a 
\quad ,\ee
where $g^\m{}_a$ is defined in eq. (\ref{gma}),
the usual procedure shows that the divergence of the Rarita-Schwinger
equation vanishes for any value of the parameter $\a$. On the other hand, the divergence
of the energy-momentum tensor presents further subtleties that we would like to 
address.  In particular, it vanishes to lowest order, 
while it gives a covariant non-vanishing result if all fermion couplings are taken into
account.  The subtleties have to do with the transformation of the vector under general coordinate transformations,
\be
\delta _\xi A_\m = - \xi^\a \de_\a A_\m - \de_\m \xi^\a A_\a
\ee
and with the corresponding (off-shell) form of the identity of eq. (\ref{emtnoether}).
Starting again from the consistent equations, one finds
\be {\cal A}_{\xi} = 2 \xi_\n D_\m T^{\m\n} +
\xi_\n tr (A^\n D_\m J^\m ) + \xi_\n tr (F^{\m\n} J_\m ) 
+ \xi_\n  (\bar{\Psi}^\n D_\m
\tilde{J}^\m )\quad. \label{emtnoetheroff}
\ee 
Reverting to the covariant form then eliminates the divergence of the Rarita-Schwinger equation
and alters the vector equation, so that the third, ``off-shell''
term, has to be retained. The final result,
\be
D_\m T^{\m\n}_{(cov)} = -\frac{1}{2} tr( A^\n D_\m J^\m_{(cov)} +
f_\m  F^{\m\n} + A^\n D_\m f^\m ) - \frac{1}{2}
e^{\n a} D_\m g^\m{}_a \ ,
\ee
is nicely verified by our equations.   
As anticipated, this implies that the divergence of $T^{\m\n}_{(cov)}$ 
vanishes to lowest order in the fermi couplings.

\section{PST construction and covariant Lagrangian formulation}

In the previous Section we have reviewed a number of properties of 6d
$(1,0)$ supergravity coupled to vector and tensor multiplets~\cite{as,fms,frs,rs}.  We have 
always confined our
attention to the field equations, thus evading the traditional difficulties met
with the action principles for (anti)self-dual tensor fields.
In this Section we would like to complete our discussion and present an action
principle for the consistent field equations. What follows is an
application of a general method introduced by
Pasti, Sorokin and Tonin (PST), that have shown how to obtain
Lorentz-covariant Lagrangians for (anti)self-dual tensors 
with a single auxiliary field~\cite{pst}. Alternative constructions~\cite{infinite},
some of which preceded the work of PST, need an infinite number of
auxiliary fields, and bear a closer relationship to the BRST formulation
of closed-string spectra~\cite{berkovits}. 
This method has already been applied to a number of systems, including 
$(1,0)$ six-dimensional supergravity 
coupled to tensor multiplets~\cite{6bpst} and type IIB 
ten-dimensional supergravity~\cite{10bpst}, whose (local) gravitational anomaly 
has been shown to reproduce~\cite{anomalypst} the 
well-known results~\cite{agw} of Alvarez-Gaum\'e and Witten.
Still, an action principle for the consistent equations reviewed in the
previous Section is of some interest since, as we have seen, these 
six-dimensional models have a number of unfamiliar properties.

Let us begin by considering a single 2-form with a self-dual  
field strength in six-dimensional Minkowski space. 
The PST lagrangian~\cite{pst}
\be
{\cal L} =\frac{1}{12}H_{\m\n\r} H^{\m\n\r} 
-\frac{1}{4(\de \phi)^2}
\de^\m \phi H^-_{\m\n\r} H^{- \s\n\r} \de_\s \phi \quad ,
\ee
where $H = d B$ and $H^- = H - *H$, is invariant under 
the gauge transformation $\delta B = d \L $,
as well as under the additional gauge transformations 
\be
\delta B = (d \phi ) \L^\prime \label{1} 
\ee 
and
\ba
& & \delta \phi = \L \quad , \nonumber \\
& & \delta B_{\m\n} = \frac{\L}{(\de \phi )^2 } H^-_{\m\n\r} \de^\r \phi \quad .
\label{2}
\ea
The last two types of gauge transformations can be used to recover
the usual field equation of a self-dual 2-form~\cite{pst}.
Indeed, the scalar equation results from the 
tensor equation contracted with
\be
\frac{H^-_{\m\n\r} \de^\r \phi }{(\de \phi )^2 } \quad ,
\ee
and consequently does not introduce any additional degrees of freedom.
The invariance of eq. (\ref{2}) can then be used to eliminate the scalar field. This
field, however, can not be set to zero, since this choice would clearly make the 
Lagrangian of eq. (\ref{1}) inconsistent. With this proviso, using eq. (\ref{1}) 
one can see that the only solution of the tensor-field equation 
is precisely the self-duality condition for its field strength.

We now want to apply this construction to six-dimensional supergravity coupled
to vector and tensor multiplets. In the conventions of ref.~\cite{frs}, 
the theory describes a single self-dual 2-form
\be
\hat{\cal{H}}_{\m\n\r} =v_r \hat{H}^r_{\m\n\r} -\frac{i}{8}(\bar{\chi}^m 
\g_{\m\n\r} \chi^m )
\ee
and $n$ antiself-dual 2-forms
\be
\hat{\cal{H}}^m_{\m\n\r} =x^m_r \hat{H}^r_{\m\n\r} +\frac{i}{4}x^m_r c^{rz}
tr_z (\bar{\l} \g_{\m\n\r} \l ) \quad .
\ee
The complete Lagrangian is obtained adding the term
\be
-\frac{1}{4}\frac{\de^\m \phi \de^\s \phi }{(\de \phi )^2 }
[ \hat{\cal H}^-_{\m\n\r} \hat{\cal H}^-_{\s}{}^{\n\r} +
\hat{\cal H}^{m+}_{\m\n\r} \hat{\cal H}^{m+}{}_\s{}^{\n\r} ] 
\ee
to eq. (2.6) of ref.~\cite{rs}.
It can be shown \cite{6bpst} that the 3-form
\be
\hat{K}_{\m\n\r} =\hat{\cal{H}}_{\m\n\r}-
3\frac{\de_{[\m} \phi \de^\s \phi }{(\de \phi )^2 }\hat{\cal{H}}^-_{\n\r ]\s}
\ee
is {\it identically} self-dual, while the 3-forms
\be
\hat{K}^m_{\m\n\r} =\hat{\cal{H}}^m_{\m\n\r}-
3\frac{\de_{[\m} \phi \de^\s \phi }{(\de \phi )^2 }\hat{\cal{H}}^{m +}_{\n\r ]\s}
\ee
are  {\it identically} antiself-dual.
With these definitions, we can display rather simply the complete 
supersymmetry transformations of the fields. Actually, only the transformations of the 
gravitino and of the tensorinos are affected, and become
\ba
& & \delta \Psi_\m =\hat{D}_\m \e +\frac{1}{4} \hat{K}_{\m\n\r}
\g^{\n\r}\e +\frac{i}{32}( \bar{\chi}^m \g_{\m\n\r} \chi^m )\g^{\n\r} \e -
\frac{3i}{8} (\bar{\e}\chi^m ) \g_\m \chi^m \nonumber\\
& & \qquad -\frac{i}{8}(\bar{\e}\g_{\m\n}
\chi^m)\g^\n \chi^m 
+\frac{i}{16} (\bar{\e}\g^{\n\r} \chi^m )\g_{\m\n\r} \chi^m -\frac{9i}{8}
v_r c^{rz} tr_z [(\bar{\e}\g_\m \l ) \l ] \nonumber\\
& & \qquad +\frac{i}{8}v_r c^{rz} tr_z [(\bar{\e}\g^\n \l ) 
\g_{\m\n} \l ]  -\frac{i}{16}v_r c^{rz} tr_z [(\bar{\e}\g_{\m\n\r} \l ) \g^{\n\r}\l ]
\quad  ,\nonumber\\
& & \delta \chi^m = \frac{i}{2} x^m_r \hat{\de_\a v^r } \g^\a \e +
\frac{i}{12} \hat{K}^m_{\a\b\g} \g^{\a\b\g}\e  +\frac{1}{2} x^m_r c^{rz} tr_z [(\bar{\e} \g_\a \l) \g^\a \l ]
\quad ,
\ea
while the scalar field $\phi$
is invariant under supersymmetry~\cite{6bpst,10bpst}.
It can be shown that the complete lagrangian transforms under supersymmetry as
dictated by the Wess-Zumino consistency conditions.

We now turn to describe the corresponding modifications of the supersymmetry algebra. 
In addition to general coordinate, gauge and supersymmetry 
transformations, the commutator of two supersymmetry 
transformations on $B^r_{\mu\nu}$ now generates 
two local PST transformations with parameters
\be
\L^\prime_{r \m} =\frac{\de^\s \phi}{(\de \phi )^2 }
(v_r \hat{\cal H}^-_{\s\m\a }-x^m_r \hat{\cal H}^{m+}_{\s\m\a} )\xi^\a 
\qquad , \quad
\L = \xi^\a \de_\a \phi \quad .\label{pst2}
\ee
The transformation of eq. (\ref{pst2}) on the scalar field $\phi$ 
is opposite to its coordinate transformation, and this gives an interpretation
of the corresponding commutator~\cite{6bpst,10bpst}
\be
[ \delta_1 , \delta_2 ] \phi =\delta_{gct}\phi +\delta_{PST}\phi =0\quad ,
\ee
that vanishes
consistently with the invariance of $\phi$ under supersymmetry.
Finally, the commutator on the vielbein determines
the parameter of the local Lorentz transformation, that is now
\ba
& & \Omega_{ab} = -\xi^\n (\w_{\n ab} -\hat{K}_{\n ab}
-\frac{i}{8}(\bar{\chi}^m \g_{\n ab }\chi^m ) )\nonumber\\
& & +\frac{1}{2}
(\bar{\chi}^m \e_1 )(\bar{\chi}^m \g_{ab}\e_2 )
-\frac{1}{2}(\bar{\chi}^m \e_2 )(\bar{\chi}^m \g_{ab}\e_1 )\nonumber\\
& & +v_r c^{rz} tr_z [(\bar{\e}_1 \g_a \l )(\bar{\e}_2 \g_b \l )
-(\bar{\e}_2 \g_a \l )(\bar{\e}_1 \g_b \l ) ]\quad .
\ea
All other parameters remain unchanged while, aside from the extension~\cite{frs}, 
the algebra closes on-shell on the modified field equations of the fermi fields.
Of course, the resulting field equations reduce to those of previous Section once
one fixes the PST gauge invariances in order to recover 
the conventional equations for (anti)self-dual tensor fields.

For completeness, we conclude by displaying
the lagrangian of six-dimensional supergravity
coupled to vector and tensor multiplets,
\ba
e^{-1}{\cal{L}} & = & -\frac{1}{4}R +\frac{1}{12}G_{rs} H^{r \m\n\r}
H^s_{\m\n\r} -\frac{1}{4} \de_\m v^r \de^\m v_r-\frac{1}{2} v_r c^{rz} tr_z
(F_{\m\n} F^{\m\n}) 
\nonumber\\ 
&- &  \frac{1}{8e}
\e^{\m\n\a\b\g\delta} c_r^z B^r_{\m\n} tr_z (F_{\a\b} F_{\g\delta}) 
 -\frac{i}{2} ( \bar{\Psi}_\m \g^{\m\n\r} D_\n [\frac{1}{2}(\w +\hat{\w} )]
\Psi_\r ) \nonumber\\
& - &\frac{i}{8}v_r [H+\hat{H}]^{r \m\n\r}(\bar{\Psi}_\m \g_\n \Psi_\r)
+\frac{i}{48} v_r [H+\hat{H} ]^r_{\a\b\g} (\bar{\Psi}_\m
\g^{\m\n\a\b\g}\Psi_\n ) \nonumber\\
& + & \frac{i}{2} ( \bar{\chi}^m \g^\m D_\m (\hat{\w})
\chi^m )  -\frac{i}{24}v_r \hat{H}^r_{\m\n\r} (\bar{\chi}^m
\g^{\m\n\r}
\chi^m ) \nonumber\\
& +& \frac{1}{4}x^m_r [\de_\n v^r +\hat{\de_\n v^r} ](\bar{\Psi}_\m \g^\n
\g^\m \chi^m)  -\frac{1}{8} x^m_r [H+\hat{H}]^{r \m\n\r} (
\bar{\Psi}_\m \g_{\n\r} \chi^m )\nonumber\\
& +& \frac{1}{24}x^m_r [H+\hat{H}]^{r \m\n\r} (\bar{\Psi}^\a \g_{\a\m\n\r}
\chi^m )+iv_r c^{rz} tr_z (\bar{\l} \g^\m
\hat{D}_\m \l )\nonumber\\
& +&
\frac{i}{2\sqrt{2}}  v_r c^{rz} tr_z [(F+\hat{F})_{\n\r}(\bar{\Psi}_\m
\g^{\n\r}\g^\m \l )] 
 \nonumber\\  & +& \frac{i}{12} x^m_r x^m_s \hat{H}^r_{\m\n\r} c^{sz} tr_z
(\bar{\l}\g^{\m\n\r} \l )- \frac{i}{2} x^m_r c^{rz} tr_z [(\bar{\chi}^m \g^\m \g^\n \l ) 
(\bar{\Psi}_\m \g_\n \l )]
\nonumber\\ 
&+&  \frac{1}{16}v_r c^{rz}tr_z (\bar{\l} \g_{\m\n\r} \l
)(\bar{\chi}^m
\g^{\m\n\r} \chi^m ) 
+\frac{1}{\sqrt{2}}x^m_r c^{rz} tr_z [(\bar{\chi}^m \g^{\m\n}
\l )\hat{F}_{\m\n} ]
\nonumber\\  
& -& \frac{i}{8}(\bar{\chi}^m
\g_{\m\n}\Psi_\r )x^m_r c^{rz} tr_z (\bar{\l} \g^{\m\n\r} \l ) 
\nonumber\\ 
& - &  \frac{3}{16}v_r c^{rz}tr_z [(\bar{\chi}^m \g_{\m\n} \l )
(\bar{\chi}^m \g^{\m\n} \l )] 
 -\frac{1}{8} v_r c^{rz} tr_z [(\bar{\chi}^m \l )(\bar{\chi}^m \l)]
\nonumber\\ 
& - & \frac{3}{4}\frac{x^m_r c^{rz}x^n_s c^{sz}}{v_t c^{tz}} tr_z
[(\bar{\chi}^m \l )(\bar{\chi}^n \l )] 
+\frac{1}{8}(\bar{\chi}^m \g^{\m\n\r} \chi^m )(\bar{\Psi}_\m \g_\n \Psi_\r )
\nonumber\\  
& + & \frac{1}{8}\frac{x^m_r c^{rz} x^n_s c^{sz}}{v_t c^{tz}} tr_z [(\bar{\chi}^m
\g_{\m\n}
\l )(\bar{\chi}^n \g^{\m\n}\l )]
-\frac{1}{8}(\bar{\chi}^m \g^\m \chi^n )(\bar{\chi}^m \g_\m
\chi^n )
\nonumber\\ & +& \frac{1}{4} (\bar{\Psi}_\m \g_\n \Psi_\r )
v_r c^{rz} tr_z(\bar{\l} \g^{\m\n\r}
\l )  -\frac{1}{2} v_r v_s c^{rz}c^{s z^\prime} tr_{z,z^\prime} [(\bar{\l}\g_\m
\l^\prime )(\bar{\l} \g^\m \l^\prime ) ]\nonumber\\
& + &  \frac{\a}{2}c^{rz} c_r^{z^\prime} tr_{z,z^\prime} [(\bar{\l}
\g_\m
\l^\prime )(\bar{\l} \g^\m \l^\prime )] \nonumber\\
& -& \frac{\de^\m \phi \de^\s \phi }{4(\de \phi )^2 }
[ \hat{\cal H}^-_{\m\n\r} \hat{\cal H}^-_{\s}{}^{\n\r} +
\hat{\cal H}^{m+}_{\m\n\r} \hat{\cal H}^{m+}{}_\s{}^{\n\r} ]
\quad,
\label{completelag}
\ea
where $\a$ is the (undetermined) coefficient of the quartic
coupling for the gauginos, 
and the corresponding supersymmetry transformations
\ba
& & \delta e_\m{}^a =-i(\bar{\e} \g^a \Psi_\m ) \quad,\nonumber\\  & &
\delta B^r_{\m\n} =i v^r (\bar{\Psi}_{[\m} \g_{\n]} \e )+ \frac{1}{2} x^{mr}
(\bar{\chi}^m
\g_{\m\n} \e )-2c^{rz} tr_z (A_{[\m}\delta A_{\n]}) \quad,
\nonumber\\  & & \delta v_r = x^m_r (\bar{\chi}^m \e )\quad,\nonumber\\  & &
\delta A_\m = -\frac{i}{\sqrt{2}} (\bar{\e} \g_\m \l ) \quad ,\nonumber\\  & &
\delta \Psi_\m =\hat{D}_\m \e +\frac{1}{4} \hat{K}_{\m\n\r}
\g^{\n\r}\e +\frac{i}{32}( \bar{\chi}^m \g_{\m\n\r} \chi^m )\g^{\n\r} \e -
\frac{3i}{8} (\bar{\e}\chi^m ) \g_\m \chi^m \nonumber\\
& & \qquad -\frac{i}{8}(\bar{\e}\g_{\m\n}
\chi^m)\g^\n \chi^m 
+\frac{i}{16} (\bar{\e}\g^{\n\r} \chi^m )\g_{\m\n\r} \chi^m -\frac{9i}{8}
v_r c^{rz} tr_z [(\bar{\e}\g_\m \l ) \l ] \nonumber\\
& & \qquad +\frac{i}{8}v_r c^{rz} tr_z [(\bar{\e}\g^\n \l ) 
\g_{\m\n} \l ]  -\frac{i}{16}v_r c^{rz} tr_z [(\bar{\e}\g_{\m\n\r} \l ) \g^{\n\r}\l ]
\quad  ,\nonumber\\
& & \delta \chi^m = \frac{i}{2} x^m_r \hat{\de_\m v^r } \g^\m \e +
\frac{i}{12} \hat{K}^m_{\m\n\r} \g^{\m\n\r}\e  +\frac{1}{2} x^m_r c^{rz} 
tr_z [(\bar{\e} \g_\m \l) \g^\m \l ]
\quad
,\nonumber\\  & & \delta \l =-\frac{1}{2\sqrt{2}}\hat{F}_{\m\n} \g^{\m\n} \e  -
\frac{1}{2} \frac{x^m_r c^{rz}}{v_s c^{sz}} (\bar{\chi}^m \l ) \e 
 - \frac{1}{4} \frac{x^m_r c^{rz}}{v_s c^{sz}} (\bar{\chi}^m \e ) \l  
\nonumber \\ & & \quad \quad + \frac{1}{8} \frac{x^m_r c^{rz}}{v_s c^{sz}}
(\bar{\chi}^m \g_{\m\n} \e ) \g^{\m\n}
\l \quad .
\ea
One further comment is in order.
Kavalov 
and Mkrtchyan \cite{infinite}
obtained long ago a complete action for pure 
d=6 (1,0) supergravity in terms of a single tensor auxiliary field.
Their work may be connected to this special case of our result 
via an ansatz relating their tensor
to the PST scalar. Still, the PST formulation has the virtue of simplicity 
and makes it manifest that the extra degrees of freedom may be locally
eliminated via additional gauge transformations\footnote{We are grateful 
to R. Mkrtchyan for explaining to us the scalar ansatz 
and to M. Tonin for related discussions.}
.
\vskip 15pt
\section*{Acknowledgments}

A.S. would like to thank the Organizers of the 1998 Trieste Conference on
``Superfivebranes and Physics in 5+1 Dimensions", the Organizers of 
the 1998 DESY Workshop
on ``Conformal Field Theory of D branes" and the Organizers of 
the 1998 Como-Centro Volta
RFBF-INTAS School on ``Advances in Quantum Field Theory, 
Statistical Mechanics and Dynamical Systems" for their kind invitation to 
present these results.

\end{document}